\documentclass[aps,reprint,prd,a4paper,portrait,preprintnumbers,nofootinbib,tightenlines,unsortedaddress,showkeys]{revtex4-1}

\usepackage{pslatex}
\usepackage{psfrag}
\usepackage{graphicx}
\usepackage{slashed}
\usepackage{xspace}
\usepackage{color}
\usepackage{amsmath}
\usepackage{amssymb}
\usepackage{placeins}

\usepackage{hyperref}
\hypersetup{bookmarksnumbered,}

\graphicspath{{figures/}{jetdistr/}}

\newcommand{\NGluon}{{\sc NGluon}\xspace}
\newcommand{\NJet}{{\sc NJet}\xspace}

\definecolor{mygreen}{rgb}{0,0.7,0}
\definecolor{mypurple}{RGB}{153,51,102}

\newcommand{\qbar}{{\overline{q}}}

\def\HThat{\ensuremath{{\widehat{H}_T}}}
\def\R#1{\ensuremath{{\cal R}_{#1}}}
\def\as{\ensuremath{\alpha_s}}
\def\mur{\ensuremath{\mu_r}}
\def\muf{\ensuremath{\mu_f}}
\def\LO{\mbox{\scriptsize LO}}
\def\NLO{\mbox{\scriptsize NLO}}
\def\njet{n\text{-jet}}
\def\parton{{\mbox{\scriptsize parton}}}
\def\shad{s_{\text{had}}}

\def\Fig#1{Fig.~\ref{#1}}
\def\Tab#1{Tab.~\ref{#1}}

\def\GeV{\text{ GeV}}

\begin{document}

\title{Next-to-leading order QCD corrections to five jet production at the LHC}

\date{\today}

\author{Simon Badger}
\affiliation{%
Niels Bohr International Academy and Discovery Center,
The Niels Bohr Institute,\\%
University of Copenhagen, Blegdamsvej 17, DK-2100 Copenhagen, Denmark}
\email{badger@nbi.dk}
\author{Benedikt Biedermann}
\affiliation{%
Humboldt-Universit\"at zu Berlin, Institut f\"ur Physik,\\%
Newtonstra{\ss}e 15, D-12489 Berlin, Germany}
\email{benedikt.biedermann@physik.hu-berlin.de}
\author{Peter Uwer}
\affiliation{%
Humboldt-Universit\"at zu Berlin, Institut f\"ur Physik,\\%
Newtonstra{\ss}e 15, D-12489 Berlin, Germany}
\email{peter.uwer@physik.hu-berlin.de}
\author{Valery Yundin}
\affiliation{%
Niels Bohr International Academy and Discovery Center,
The Niels Bohr Institute,\\%
University of Copenhagen, Blegdamsvej 17, DK-2100 Copenhagen, Denmark}
\email{yundin@nbi.dk}
\author{}
\affiliation{}

\preprint{HU-EP-13/48}
\preprint{SFB/CPP-13-67}

\begin{abstract}
  We present theoretical predictions for five jet production in proton-proton collisions at
  next-to-leading order accuracy in QCD. Inclusive as well as differential observables are studied
  for collision energies of 7 and 8~TeV. In general the next-to-leading order corrections stabilize
  the theoretical predictions with respect to scale variations. In case of the inclusive jet cross
  sections, we compare with experimental data where possible and find reasonable agreement. We
  observe that the four-to-three and five-to-four jet ratios show better perturbative convergence
  than the known three-to-two ratio and are promising candidates for future $\alpha_s$ measurements.
  Furthermore, we present a detailed analysis of uncertainties related to parton distribution
  functions. The full colour virtual matrix elements used in the computation were obtained with the
  \NJet package \cite{Badger:2012pg}, a publicly available library for the evaluation of one-loop
  amplitudes in massless QCD.
\end{abstract}



\maketitle

\section{Introduction}

The wealth of data recorded at the LHC experiments in run~1 presents an excellent opportunity to
test quantum chromo-dynamics (QCD) in a new energy regime. Furthermore, multi-jet production at
large transverse momentum can provide useful information to constrain the parton distribution
functions. Where precise theoretical results are available, the data can also be used to determine
the strong coupling constant $\as$. In that context large multiplicities may turn out to be
particularly useful, in analogy to what has been observed in electron-positron annihilation. Pure
QCD reactions with multiple jet production can also give large backgrounds to various new physics
searches therefore precision predictions are required.

Next-to-leading order (NLO) predictions at fixed order in $\as$ for di-jet production have been
known for more than 20~years
\cite{Giele:1993dj}.  Three-jet production represents a considerable increase in computational
complexity and a full computation was completed in 2002 \cite{Nagy:2001fj}, and implemented in the
public code {\sc NLOJET++}, though pure gluonic contributions were known previously
\cite{Kilgore:1996sq}.  Breakthroughs in virtual amplitude computations have recently enabled
predictions of four-jet production \cite{Bern:2011ep,Badger:2012pf}, with results generally in good
agreement with the experimental data \cite{Aad:2011tqa}. Di-jet production is known to suffer from
large corrections from soft gluon radiation which requires resummation beyond fixed order
perturbation theory. Theoretical predictions at NLO including the parton shower (NLO+PS) allow to
account for these effects and obtain a better description of the available data
\cite{Alioli:2010xa,Hoeche:2012fm}. The CMS collaboration has recently completed a measurement of
$\alpha_s$ from the inclusive 3-jet to inclusive 2-jet ratio. The measurement illustrates well the
LHC potential for precision measurements. With NNLO QCD corrections to di-jet production just around
the corner \cite{Ridder:2013mf}, theoretical uncertainties look to be under good control and enable
future QCD predictions to move beyond the `industry standard' LO+PS/ME+PS accuracy.

By now there has been a great success of the program to automate NLO
computations. Processes with four or five final state particles, previously thought to be
impossible, are now available in public codes
\cite{Badger:2010nx,Hirschi:2011pa,Bevilacqua:2011xh,Cullen:2011ac,Badger:2012pg}.
New methods and algorithms \cite{Cascioli:2011va,Becker:2011vg,Actis:2012qn} have been pushing the boundaries of
perturbative QCD computations and increasing the range of phenomenological applications (recent
examples can be found in
\cite{Bevilacqua:2012em,Bern:2013gka,Cullen:2013saa,vanDeurzen:2013xla,Gehrmann:2013bga},
see also
\cite{AlcarazMaestre:2012vp} for a more complete overview). Powerful on-shell unitarity methods
have been established for many years \cite{Bern:1994zx,Bern:1994cg} yet continue to demonstrate
the ability to simplify calculations of extremely high multiplicity
amplitudes inaccessible with alternative tools. Recent state-of-the-art computations include
the NLO QCD corrections to $pp\to W+5$ jets \cite{Bern:2013gka} by the {\sc BlackHat} collaboration. In
this paper we present the first computation for the production of five hard jets at NLO in QCD.

The article is organized as follows: we first give an overview of the computational framework for the
various partonic sub-channels in Section \ref{sec:calc}. Section \ref{sec:5j} contains our results
for the NLO QCD corrections to $pp \to 5$ jets. We describe the numerical setup and interface with
the Sherpa Monte-Carlo event generator \cite{Gleisberg:2008ta} in Section \ref{sec:numsetup} along
with kinematic cuts. Results for total cross sections and jet ratios are compared to the available
data in Section \ref{sec:results}. We present differential distributions in the jet transverse
momenta and rapidity and perform a detailed comparison of different PDF fits. We finally present our
conclusions and outlook for the future. Two appendices containing a complete set of distributions
and numerical values used in the plotted histograms are included to ease future comparisons.

\section{Outline of the calculation \label{sec:calc}}

The calculation is done in QCD with five massless quark flavours including the bottom-quark in the
initial state. We expect the neglected contributions from top quark loops to be much smaller than
the estimated theoretical uncertainty. The processes contributing to five-jet production may be
derived from the following four basic channels via crossing symmetry
\begin{align}
  0&\to g g  g g g g g,& 0&\to q \qbar g g g g g, \nonumber\\
  0&\to q \qbar q' \qbar'  g g g,&  0&\to q \qbar q' \qbar' q'' \qbar'' g\nonumber
\end{align}
where $q$, $q'$ and $q''$ denote generic quarks of different flavour.
Amplitudes with like-flavour quark pairs can always be obtained from
the amplitudes for different flavours by an appropriate
(anti) symmetrization.  The $n$-jet
differential cross section expanded in the coupling $\as$ reads
\begin{equation}
  d\sigma_n = d\sigma_n^{\LO} +  d\delta\sigma_n^{\NLO}
  + {\cal O}(\as^{n+2})
\end{equation}
where $d\sigma_n^{\LO}\sim \as^n$ and $d\delta\sigma_n^{\NLO} \sim
\as^{n+1}$. In the QCD improved parton model, the leading order
differential cross section $d\sigma_n^{\LO}$ is given by
\begin{align}
  d\sigma_n^{\LO} =& \sum_{\substack{i,j\\ \in \{q,\qbar,g\}}} dx_i dx_j
  F_{j/H_2}(x_j,\muf)F_{i/H_1}(x_i,\muf)
  \nonumber \\
  &
  {}\times
  d\sigma_n^{\rm B}\bigl(i(x_iP_1)+j(x_jP_2) \to n \mbox{ part.}\bigr).
  \label{eq:HadXSection}
\end{align}
$P_1, P_2$ are the momenta of the two incoming hadrons $H_1,H_2$ which we assume to be massless.
The total incoming momentum of the initial state partons $P=x_i P_1 + x_j P_2$ leads to a partonic centre-of-mass energy squared $\hat s = 2 x_i x_j
(P_1\cdot P_2)=x_1x_2s_{\rm had}$ with $s_{\rm had}$ being the hadronic centre-of-mass energy
squared. The parton distribution functions $F_{i/H}(x,\muf)$ describe, roughly speaking, the
probability to find a parton $i$ inside a hadron $H$ with a momentum fraction
between $x$ and $x + dx$. Note that the parton distribution functions depend besides $x$
also on the unphysical factorization scale \muf. The partonic differential cross section
$d\sigma_n^B\bigl(i(x_1P_1)+j(x_jP_2) \to n\mbox{ part.}\bigr)$ describes the reaction $(ij \to
n\mbox{-jets})$ in Born approximation using the leading order matrix elements $|{\cal M}_n(ij\to
n\mbox{ part.})|^2$ and a suitable jet algorithm $\Theta_{\njet}$:
\begin{multline}
  d\sigma_n^{\rm B} =
  \frac{1}{2 \hat s}\prod_{\ell =1}^n\frac{d^3k_\ell }{ (2\pi)^3 2E_\ell}
  \Theta_{\njet}
  \\
  {}\times(2\pi)^4
  \delta\Bigl(P-\sum_{m=1}^n k_m\Bigr)\big|{\cal M}_n(ij\to n \mbox{ part.})\big|^2.
  \label{eq:PartonXSection}
\end{multline}
$k_i$ are the four momenta of the outgoing partons obeying $\hat s = (\sum_{i=1}^n k_i)^2$. The jet
algorithm $\Theta_{\njet}$ is a function depending solely on the final state parton momenta $k_i$
and on parameters defining the geometric extensions of the jet. Its value is equal to one
if the final state momentum configuration corresponds to a valid $n$-jet event and zero otherwise.
The Born matrix elements ${\cal M}(ij\to n \mbox{ part.})$ have been evaluated with Comix
\cite{Gleisberg:2008fv} within the Sherpa framework. The Sherpa Monte Carlo event generator
\cite{Gleisberg:2007md} has also been used to perform the numerical phase space integration.

At NLO accuracy both the virtual corrections $d\sigma_n^{\rm V}$ (one-loop contribution interfered
with the Born amplitude) and the real corrections $d\sigma_{n+1}^{\rm R}$ (tree-level
amplitudes with one additional parton in the final state) contribute to the $n$-jet cross section.
Both $d\sigma_n^{\rm V}$ and $d\sigma_{n+1}^{\rm R}$ contain separately collinear and soft
divergences. Only after combining the two contributions and factorizing the
initial state singularities into renormalized parton distributions one obtains a finite result. In
order to perform the cancellation of the divergences numerically we apply the Catani-Seymour
subtraction method \cite{Catani:1996jh}.  The idea is to add and subtract local
counter-terms $d\sigma^{S}_{n+1}$ with $(n+1)$-parton kinematics which, on the one hand, mimic
point-wise the singularity structure of the real corrections and which, on the other hand, are
chosen such that the singularity due to the additional parton emission can be calculated
analytically via separate integration of the one-particle phase-space. Since the latter cancel by
construction the divergences from the virtual corrections, only finite quantities remain making the
numerical integration with a Monte Carlo program feasible. Schematically, we may write the total
cross section as
\begin{multline}
  \delta\sigma^{\NLO} = \int\limits_n
  \big(d\sigma_n^{\rm V}
  + \int\limits_1 d\sigma_{n+1}^{\rm S}\big)
  + \int\limits_n d\sigma_n^{\rm Fac}
    \\
  + \int\limits_{n+1} \big(d\sigma_{n+1}^{\rm R} - d\sigma_{n+1}^{\rm S}\big).
\end{multline}
$d\sigma_n^{\rm Fac}$ is due to the
factorization of initial state singularities.
The NLO corrections can then be written in terms of three finite contributions:
\begin{align}
  d\delta\sigma_n^{\NLO} &= d\bar\sigma_n^{\rm V} + d\bar\sigma_n^{\rm
    I} + d\sigma_{n+1}^{\rm RS},
  \label{eq:NLOxs}
\end{align}
$d\bar\sigma_n^{\rm V}$ denotes the finite part of the virtual corrections, $d\bar\sigma_n^{\rm I}$
the finite part of the integrated subtraction terms together with the contribution from the
factorization and $d\sigma_{n+1}^{\rm RS}$ the real corrections combined with the subtraction terms.
For the computation of $d\bar\sigma_n^{\rm I}$ and $d\sigma_n^{\rm RS}$ we use Sherpa which provides
a numerical implementation of the Catani-Seymour subtraction scheme. The required tree-level
amplitudes are, as in the LO case, computed with Comix as part of the Sherpa framework.

The necessary one-loop matrix elements for the virtual corrections $d\bar\sigma_n^{\rm V}$ are
evaluated with the publicly available \NJet\footnote{To download \NJet visit the project home page
at\\ \url{https://bitbucket.org/njet/njet/}.} package \cite{Badger:2012pg}. \NJet uses an on-shell
generalized unitarity framework \cite{Britto:2004nc,Ellis:2007br,Forde:2007mi,Berger:2008sj} to
compute multi-parton one-loop primitive amplitudes from tree-level building blocks. An accurate
numerical implementation is achieved using the integrand reduction procedure of OPP
\cite{Ossola:2006us}. The algorithm is based on the \NGluon library \cite{Badger:2010nx} following
the description of $D$-dimensional generalized unitarity presented in Refs.
\cite{Giele:2008ve,Badger:2008cm} and using Berends-Giele recursion \cite{Berends:1987me} for
efficient numerical evaluation of tree-level amplitudes.  For a more detailed description of the employed
methods and the usage of the program, we refer to Refs. \cite{Badger:2010nx,Badger:2012pg}.  The
scalar loop integrals are obtained via the {\sc QCDLoop/FF package}
\cite{vanOldenborgh:1990yc,Ellis:2007qk}.  We note that \NJet is so far the only publicly available
tool that is able to compute all one-loop seven-point matrix-elements that contribute to five-jet
production in hadronic collisions. For reference numerical evaluations of the one-loop matrix
elements at a single phase-space point have been presented previously \cite{Badger:2012pg}.

\section{Results for 5-Jet production at the LHC at 7 and 8~TeV \label{sec:5j}}

\subsection{Numerical setup \label{sec:numsetup}}

As mentioned earlier we use
the Sherpa Monte-Carlo event generator \cite{Gleisberg:2008ta} to handle phase-space integration and
generation of tree-level and Catani-Seymour dipole subtraction terms using the colour dressed
formalism implemented in Comix \cite{Gleisberg:2008fv,Gleisberg:2007md}. The virtual matrix elements
are interfaced using the Binoth Les Houches Accord \cite{Binoth:2010xt,Alioli:2013nda}.

To combine partons into jets we use the anti-kt jet clustering algorithm as implemented
in {\sc FastJet} \cite{Cacciari:2011ma,Cacciari:2008gp}. Furthermore asymmetric cuts on the jets ordered
in transverse momenta, $p_T$, are applied to match the ATLAS multi-jet measurements \cite{Aad:2011tqa}:
\begin{align}
  p_T^{j_1} &> 80 \GeV, & p_T^{j_{\geq 2}} &> 60 \GeV, & R &= 0.4.
  \label{eq:cuts}
\end{align}

The PDFs are accessed through the LHAPDF interface \cite{Whalley:2005nh} with all central values using
NNPDF2.1~\cite{Ball:2011uy} for LO ($\alpha_s(M_Z) = 0.119$) and NNPDF2.3~\cite{Ball:2012cx} for NLO
($\alpha_s(M_Z) = 0.118$) if not mentioned otherwise.

Generated events are stored in Root Ntuple format \cite{Binoth:2010ra} which allows for flexible
analysis.  Renormalization and factorization dependence can be re-weighted at the analysis level as
well as the choice of PDF set. Since the event generation of high multiplicity processes at NLO is
computationally intensive analysis of PDF uncertainties and scale choices
 would be prohibitive without this technique.

\subsection{Numerical results \label{sec:results}}

In this section we present the numerical results for total cross sections and selected\footnote{The complete set of results presented in this
section together with additional distributions for 7 and 8 TeV can be obtained from
\url{https://bitbucket.org/njet/njet/wiki/Results/Physics}.} distributions
at centre-of-mass energies of 7 and 8 TeV. Within the setup described in the
previous section we have chosen the renormalization and factorization scales to be equal
$\mur=\muf=\mu$ and use a dynamical scale based on the total transverse momentum $\HThat$ of the
final state partons:
\begin{equation}
  \HThat = \sum_{i=1}^{N_\parton} p_{T,i}^\parton.
\end{equation}
We then obtain the 5-jet cross section at 7~TeV,
\begin{center}
  \setlength{\tabcolsep}{12pt}
  \renewcommand{\arraystretch}{1.3}
  \begin{tabular}{l|c|c}
    $\mu$ & $\sigma_5^{\text{7TeV-LO}} \: [{\rm nb}]$ & $\sigma_5^{\text{7TeV-NLO}} \: [{\rm nb}]$ \\
    \hline
    $\HThat/2$ & $0.699 ( 0.004 )$ & $0.544 ( 0.016 )$ \\
    $\HThat$   & $0.419 ( 0.002 )$ & $0.479 ( 0.008 )$ \\
    $\HThat/4$ & $1.228 ( 0.006 )$ & $0.367 ( 0.032 )$
  \end{tabular}
\end{center}
where numerical integration errors are quoted in parentheses. We show the values of the cross section
at three values of the renormalization scale, $\mu=x\HThat/2$ where $x = 0.5, 1, 2$. We observe
significant reduction in the residual scale dependence when including NLO corrections. Within the
chosen scale band, the LO predictions lie within a range of $0.810$ nb while at NLO the range is
$0.177$ nb. The analagous results at 8~TeV are shown below.
\begin{center}
  \setlength{\tabcolsep}{12pt}
  \renewcommand{\arraystretch}{1.3}
  \begin{tabular}{l|c|c}
    $\mu$ & $\sigma_5^{\text{8TeV-LO}} \: [{\rm nb}]$ & $\sigma_5^{\text{8TeV-NLO}} \: [{\rm nb}]$ \\
    \hline
    $\HThat/2$ & $1.044 ( 0.006 )$ & $0.790 ( 0.021 )$ \\
    $\HThat$   & $0.631 ( 0.004 )$ & $0.723 ( 0.011 )$ \\
    $\HThat/4$ & $1.814 ( 0.010 )$ & $0.477 ( 0.042 )$
  \end{tabular}
\end{center}
\begin{figure}[htbp]
  \begin{center}
    \leavevmode
    \includegraphics[width=1.1\columnwidth]{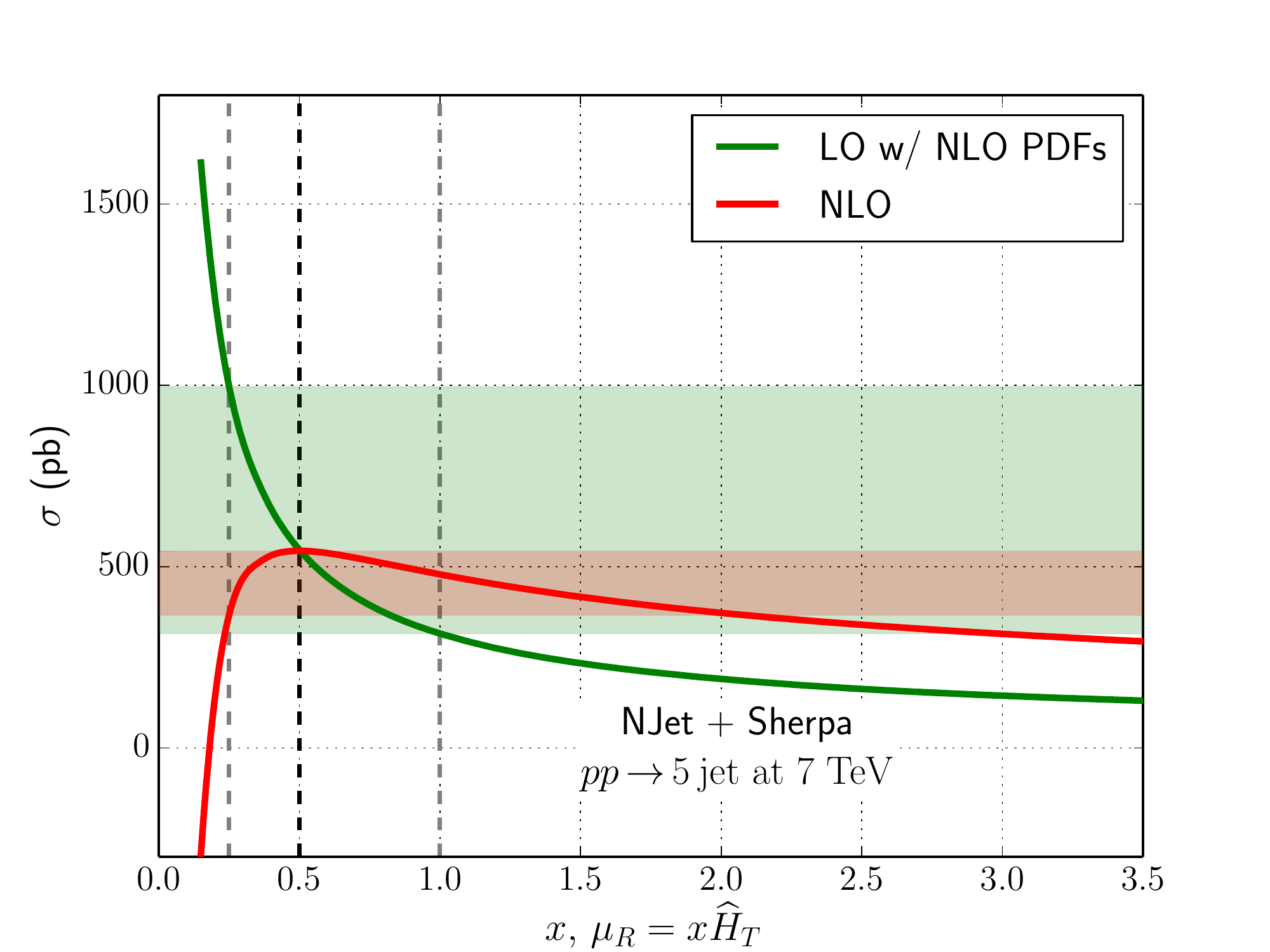}
    \caption{Same as \Fig{fig:5j_scalevar} but using the NLO setup in LO.}
    \label{fig:5j_scalevar_nlopdfs}
  \end{center}
\end{figure}
In \Fig{fig:5j_scalevar_nlopdfs} the scale dependence of the leading order and next-to-leading order
cross section is illustrated. The dashed black line indicates $\mu=\HThat/2$. The horizontal bands show
the variation of the cross section for a scale variation between $\HThat/4$ and $\HThat$. The
uncertainty due to scale variation is roughly reduced by a factor of one third. Furthermore we see
that around $\mu=\HThat/2$ the NLO cross section is flat indicating that $\mu=\HThat/2$ is a
reasonable choice for the central scale.  This is further supported by the fact that for
$\mu=\HThat/2$ the NLO corrections are very small. It is also interesting to observe that the two bands,
LO and NLO, nicely overlap. Note however that we have used the NLO setup in the leading order
calculation. In particular the NLO PDFs with the corresponding $\as$ are employed.
\begin{figure}[htbp]
  \begin{center}
    \leavevmode
    \includegraphics[width=1.1\columnwidth]{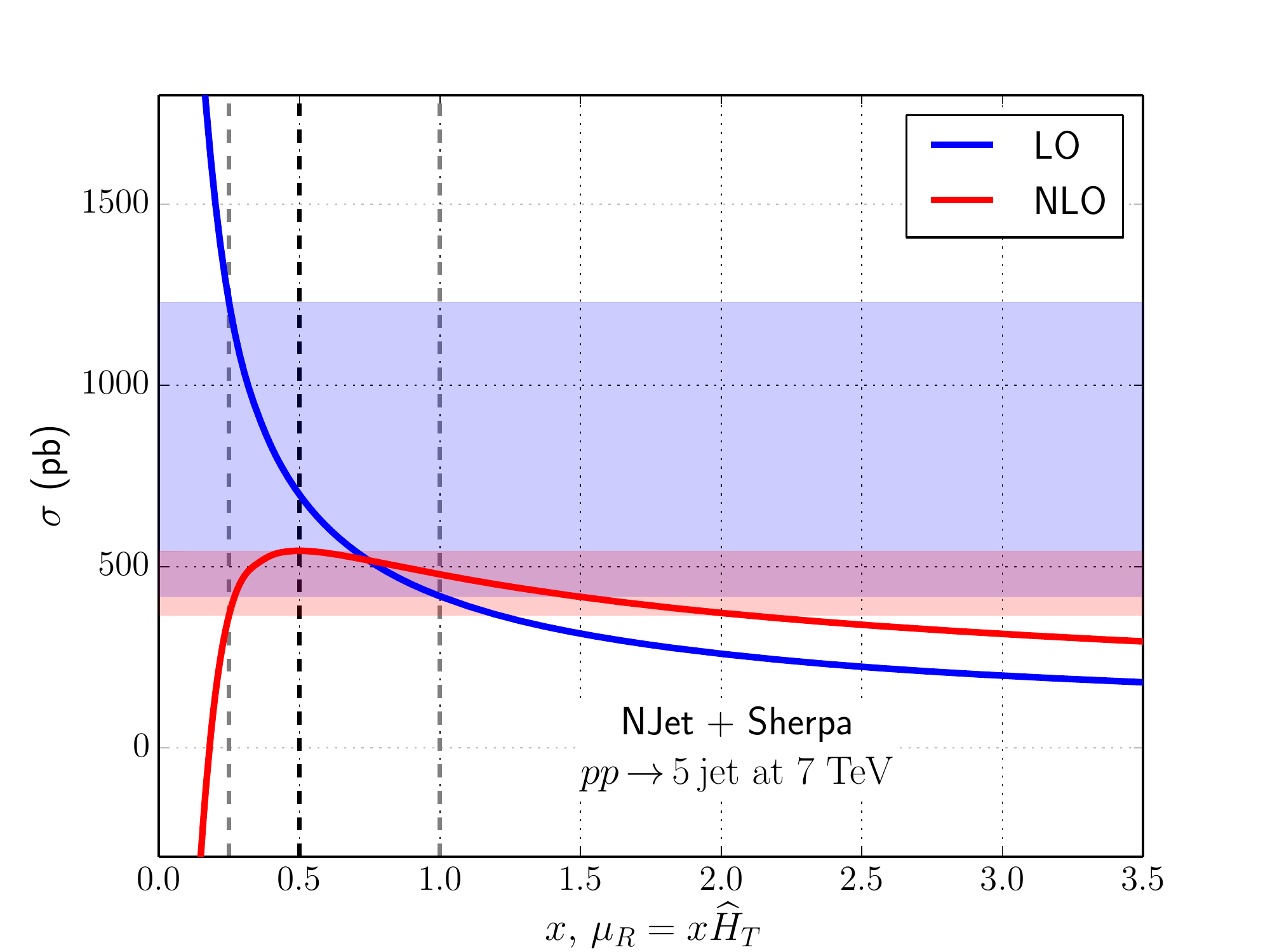}
    \caption{Residual scale dependence of the 5-jet cross section in
    leading and next-to-leading order.}
    \label{fig:5j_scalevar}
  \end{center}
\end{figure}
In \Fig{fig:5j_scalevar} we show the scale dependence using in the leading order prediction LO PDFs
with the respective $\as$.
Compared to \Fig{fig:5j_scalevar_nlopdfs} we observe in
\Fig{fig:5j_scalevar} a much larger difference between the LO and NLO
prediction. To some extend the difference is due to the change in
\as. Similar to what has been found in Ref.~\cite{Badger:2012pf} we
conclude that using the NLO PDFs in the LO predictions gives a
better approximation to the full result compared to using LO PDFs.

Although not a physical observable it is interesting to ask how the
different partonic channels contribute to the inclusive 5-jet rate. Ignoring different quark flavours
we distinguish nine partonic channels in LO:
\begin{eqnarray*}
  &&gg\to5g,\;gg\to qq+3g,\; qg\to q+4g, \\
  &&qq \to 5g,\; gg \to 4q+ g,\;
  qg\to 3q+2g, \\
  &&qq\to qq+3g,\; qg\to 5q,\; qq\to 4q+g,
\end{eqnarray*}
where $q$ may be any quark or anti-quark with exception of the
top-quark i.e. $qq=\{uu,u\bar u, ud, u\bar d, \ldots\}$.
In \Tab{tab:channels} the individual contribution of each channel is
presented.
\begin{table}[htbp]
  \caption{\label{tab:channels} Contribution of individual partonic channels.}
  \begin{center}
\setlength{\tabcolsep}{10pt}
    \begin{tabular}{|l|r@{.}l|}
      \hline
      $qg\to q  + 4g$ & $39$&$2\%$ \\
      $gg\to 5g $     & $27$&$3\%$ \\
      $qq\to 2q + 3g$ & $13$&$5\%$ \\
      $qg\to 3q + 2g$ & $9$&$0\%$ \\
      $gg\to 2q + 3g$ & $8$&$5\%$ \\
      $qq\to 4q + g$  & $1$&$8\%$ \\
      $gg\to 4q + g$  & $0$&$5\%$ \\
      $qg\to 5q$      & $0$&$2\%$ \\
      $qq\to 5g$      & $0$&$04\%$ \\
      \hline
    \end{tabular}
  \end{center}
\end{table}
The most important contribution is provided by the $qg$ initial state.  Almost 50\% of the cross
section can be attributed to this channel.  This is a consequence of the large parton luminosity in
combination with the sizeable cross sections. Among the $qg$ initiated reactions the $qg\to q+4g$
channel is with about 40\% of the cross section the most important process. Replacing the quark line
in this process by a gluon will still lead to large partonic cross sections. However the $gg$ parton
flux is reduced compared to the $qg$ initial state. As a consequence the purely gluonic reaction
leads to a slightly smaller contribution and is responsible for about 25\% of the cross section. The
composition of the cross section may provide useful information when jet rates are used to constrain
the PDFs. Since the luminosity functions
\begin{equation}
  {\cal L}_{ij}(\hat s, \shad,\muf) =
  {1\over \shad}\int\limits_{\hat s}^{\shad}{ds\over s}
  F_{i/p} \Big(\muf,{s\over\shad}\Big)F_{j/p}\Big(\muf,{\hat s\over s}\Big)
\end{equation}
depend on the partonic centre-of-mass energy, the composition may be
different for different kinematical configurations. We come back to
this point when we discuss differential distributions.

\begin{figure}[h]
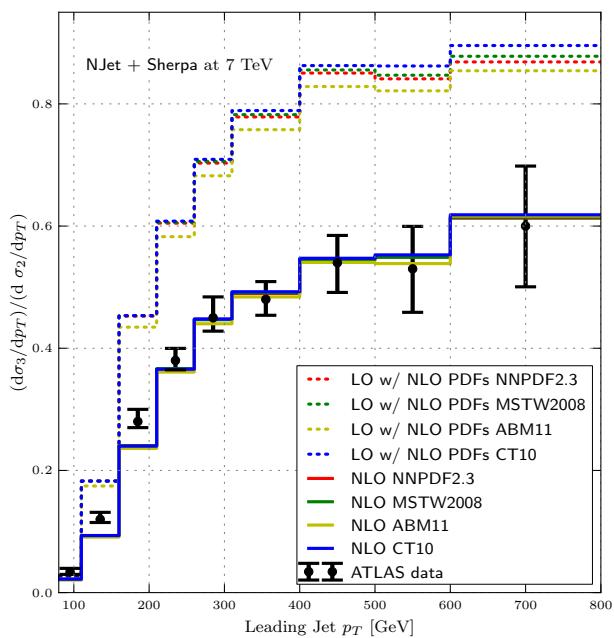

  \begin{center}
    \includegraphics[width=1.\columnwidth]{{{plot_32ratio_data_pT_1}}}
  \end{center}
  \caption{The 3/2 jet ratio as a function of the $p_T$ of the leading jet. ATLAS
  data is taken from \cite{Aad:2011tqa}. The cuts are given in section \ref{sec:numsetup}
  except the jet cone radius which is taken as $R=0.6$.}
  \label{fig:32ratiodata}
\end{figure}

\begin{figure}[htbp]
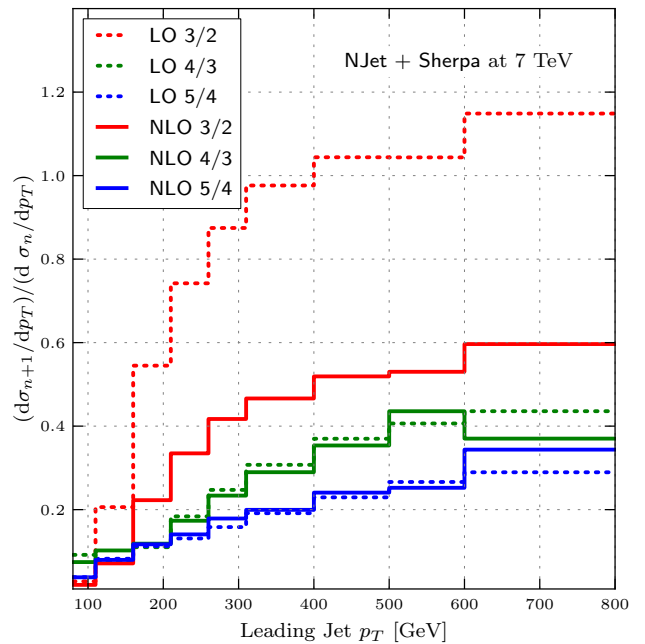

  \begin{center}
    \leavevmode
    \includegraphics[width=1.\columnwidth]{{{plot_jetratio_pT_1}}}
    \caption{The $\R{n}$ ratio as a function of the $p_T$ of the leading jet.}
    \label{fig:jetratio_pT}
  \end{center}
\end{figure}

In \Tab{tab:JetXS} we show for completeness the cross sections
for two, three and four-jet production as calculated with \NJet using
the same setup as in the five jet case.
\begin{table}[htbp]
  \renewcommand{\arraystretch}{1.3}
  \begin{center}
    \setlength{\tabcolsep}{6pt}
    \begin{tabular}{|l|c|c|c|}
      \hline
      $\mu$ & $\sigma_2^{\text{7TeV-NLO}} \: [{\rm nb}]$
            & $\sigma_3^{\text{7TeV-NLO}} \: [{\rm nb}]$
            & $\sigma_4^{\text{7TeV-NLO}} \: [{\rm nb}]$  \\
      \hline
      $\HThat/2$ & $1175 ( 3 )$ &%
                   $52.5 ( 0.3 )$ &%
                   $5.65 ( 0.07 )$ \\
      $\HThat$   & $1046 ( 2 )$ &%
                   $54.4 ( 0.2 )$ &%
                   $5.36 ( 0.04 )$ \\%
      $\HThat/4$ & $1295 ( 4 )$ &%
                   $33.2 ( 0.4 )$ &%
                   $3.72 ( 0.12 )$ \\%
      \hline
    \end{tabular}
    \caption{Results for two, three and four-jet production with the same setup as in the five-jet
    case. All values in units of nb.}
    \label{tab:JetXS}
  \end{center}
\end{table}
The real corrections to five-jet production allow us to calculate also the cross section for six jet
production, however only in leading order QCD. The result is given by
\begin{center}
  \setlength{\tabcolsep}{12pt}
  \renewcommand{\arraystretch}{1.3}
  \begin{tabular}{l|c|c}
    $\mu$ & $\sigma_6^{\text{7TeV-LO}} \: [{\rm nb}]$ & $\sigma_6^{\text{8TeV-LO}} \: [{\rm nb}]$ \\
    \hline
    $\HThat/2$ & $0.0496 ( 0.0005 )$ & $0.0844 ( 0.0010 )$ \\
    $\HThat$   & $0.0263 ( 0.0003 )$ & $0.0452 ( 0.0005 )$ \\
    $\HThat/4$ & $0.0992 ( 0.0011 )$ & $0.1673 ( 0.0021 )$
  \end{tabular}
\end{center}
where the NNPDF2.3 NLO PDF set with $\as=0.118$ has been used.  The jet rates have been measured
recently by ATLAS using the 7~TeV data set~\cite{Aad:2011tqa}.
\begin{figure}[htbp]
  \begin{center}
    \leavevmode
    \includegraphics[width=1.\columnwidth]{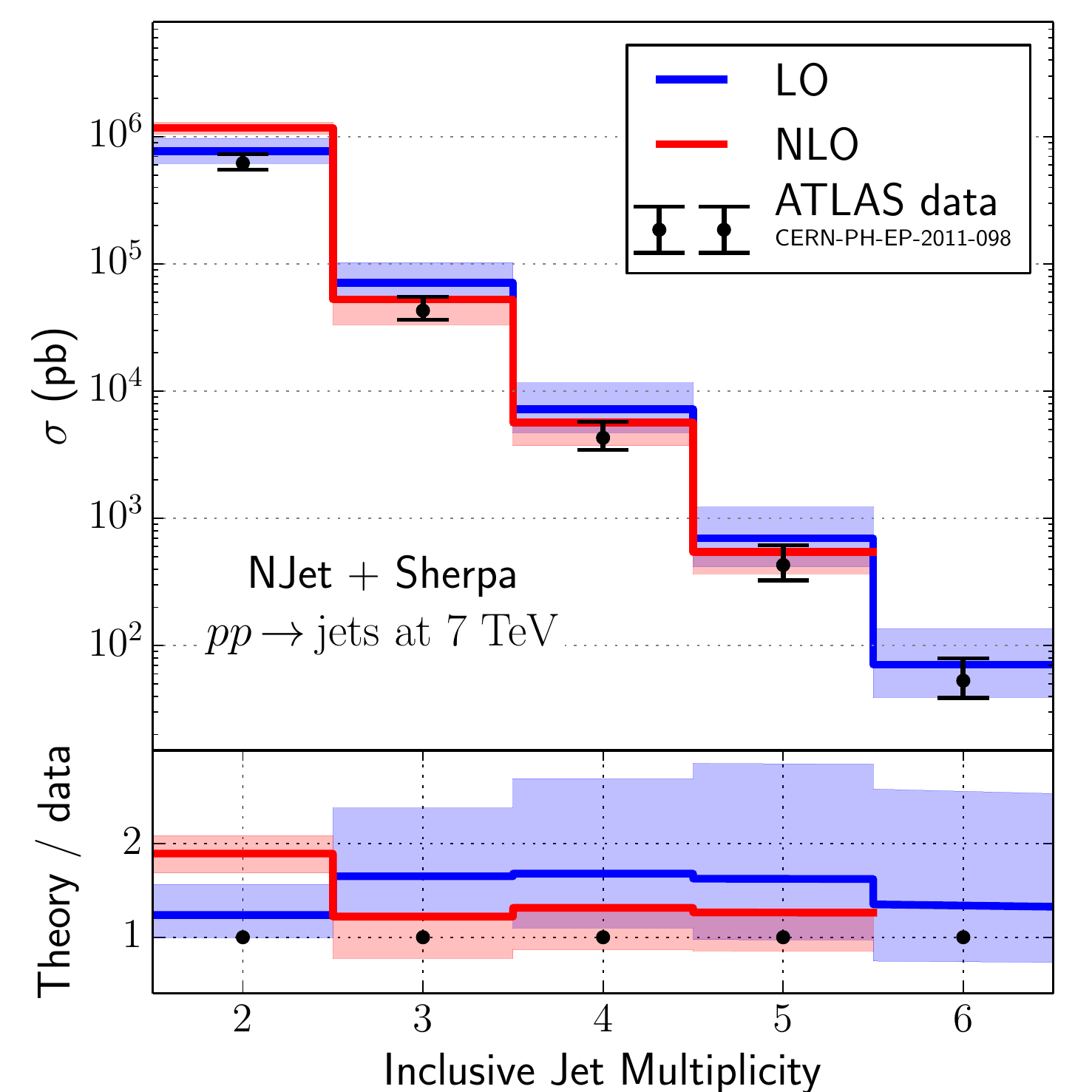}
    \caption{Cross sections for 2-, 3-, 4-, 5- and 6-jet production in
    leading and next-to-leading order as calculated with \NJet as well
    as results from ATLAS measurements \cite{Aad:2011tqa}. All LO quantities
    use NNPDF2.1 with $\alpha_s(M_Z) = 0.119$. NLO quantities use NNPDF2.3 with $\alpha_s(M_Z) = 0.118$,
    the 6-jet cross section is only avaiable LO accuracy.
    }
    \label{fig:MJincl}
  \end{center}
\end{figure}
In \Fig{fig:MJincl} we show the data together with the theoretical predictions in leading and
next-to-leading order. In case of the six jet rate only LO results are shown. In the lower plot the
ratio of theoretical predictions with respect to data is given. With exception of the two jet cross
section the inclusion of the NLO results improves significantly the comparison with data. For the
higher multiplicities where NLO predictions are available the ratio between theory and data is about
$1.2-1.3$. Given that inclusive cross sections are intrinsically difficult to measure we
consider this agreement as remarkable good. In particular for three-, four- and five-jet production the
theoretical predictions agree within the uncertainties with the data. One should also keep in mind
that a one per cent uncertainty of the collider energy may lead to sizeable changes in the cross
sections. (For example, the inclusive cross section for top-quark pair production changes by about
3\% when the energy is changed from 7~TeV to (7$\pm$0.07)~TeV.)  Instead of studying inclusive cross
sections it is useful to consider their ratios since many theoretical
and experimental uncertainties (i.e. uncertainties due to luminosity, scale dependence, PDF
dependence etc.) may cancel between numerator
and denominator.  In particular one may consider
\begin{equation}
  \R{n} =  {\sigma_{(n+1)\text{-jet}}\over\sigma_{\njet}}.
\end{equation}
This quantity is in leading order proportional to the QCD coupling $\as$ and can be used to
determine the value of $\as$ from jet rates.
\begin{figure}[htbp]
  \begin{center}
    \leavevmode
    \includegraphics[width=1.\columnwidth]{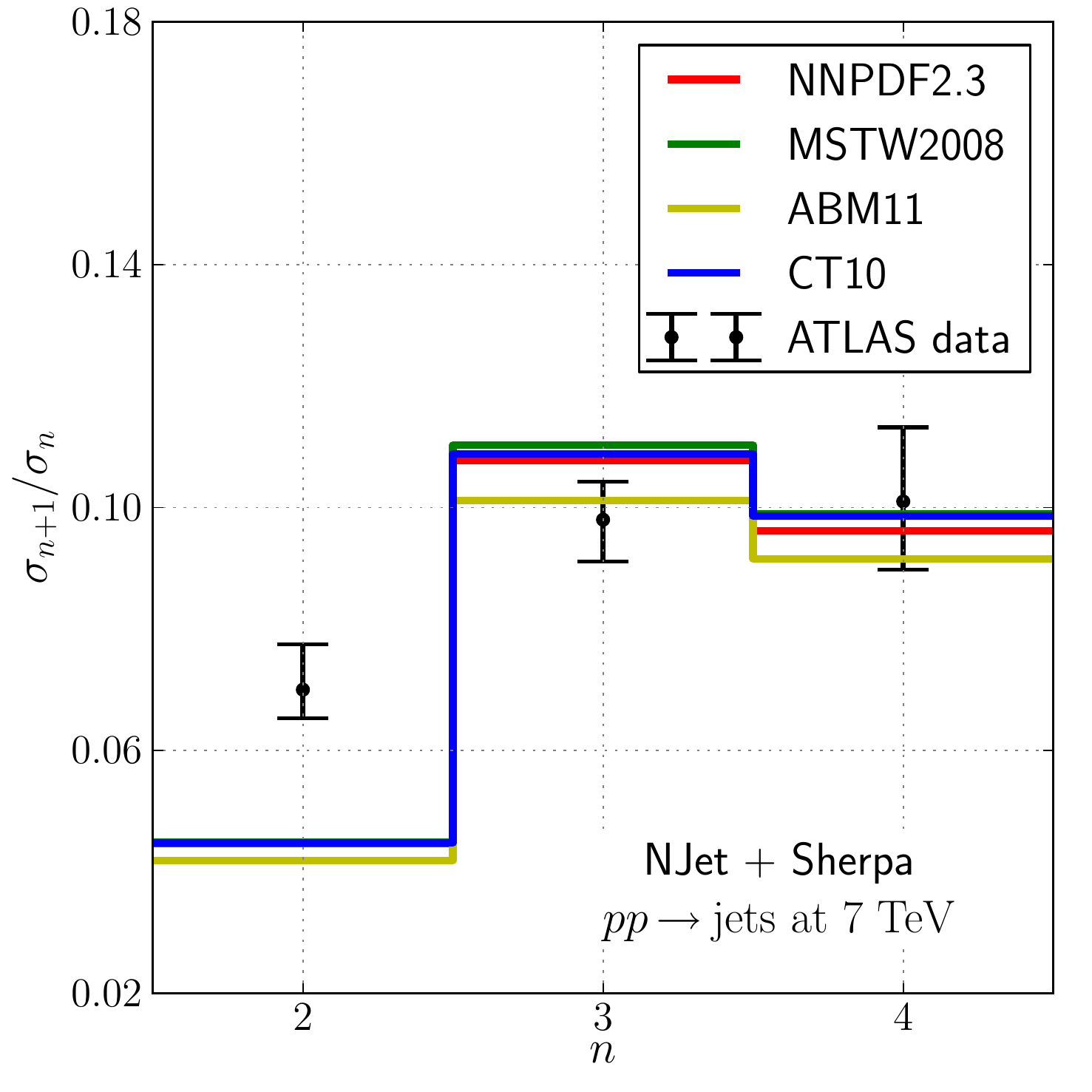}
    \caption{Theoretical predictions for the jet ratios ${\cal{R}}_n$ compared
    with recent ATLAS measurements \cite{Aad:2011tqa}. Theoretical predictions
    are made with the central values of the 4 listed PDF sets with NLO $\alpha_s$
    running. $\alpha_s(m_Z) = 0.118$ for NNPDF2.3, CT10 and ABM11 and $\alpha_s(m_Z) = 0.120$ for
    MSTW2008}
    \label{fig:jetratios}
  \end{center}
\end{figure}
In \Fig{fig:jetratios} we show QCD predictions in NLO using different PDF sets together with the
results from ATLAS. The results obtained from NNPDF2.3 are also collected in \Tab{tab:jetratios}
where, in addition, the ratios at leading order (using the LO setup with NNPDF2.1) are shown.
\begin{table}[htbp]\renewcommand{\arraystretch}{1.6}
  \begin{center}
    \begin{tabular}{|c|c|c|c|}
      \hline
      \R{n} & ATLAS\cite{Aad:2011tqa} & LO & NLO \\ \hline
      2 & $0.070^{+ 0.007 }_{- 0.005 }$ & $0.0925(0.0002)$  & $0.0447(0.0003)$\\ \hline
      3 & $0.098^{+ 0.006 }_{- 0.007 }$ & $0.102(0.000)$  & $0.108(0.002)$\\ \hline
      4 & $0.101^{+ 0.012 }_{- 0.011 }$ & $0.097(0.001)$  & $0.096(0.003)$\\ \hline
      5 & $0.123^{+ 0.028 }_{- 0.027 }$ & $0.102(0.001)$  & $--$\\ \hline
    \end{tabular}
    \caption{Results for the jet ratios $\R{n}$ for the central scale of $\HThat/2$ and NNPDF2.3 PDF
    set.}
    \label{tab:jetratios}
  \end{center}
\end{table}
In case of \R{3}\ and \R{4}\ perturbation theory seems to provide stable results. The leading order
and next-to-leading order values differ by less than 10\%. In addition NNPDF~\cite{Ball:2012cx},
CT10~\cite{Lai:2010vv} and MSTW08~\cite{Martin:2009iq} give compatible predictions.
ABM11~\cite{Alekhin:2012ig} gives slightly smaller results for \R{3}\ and \R{4}.
Within uncertainties the predictions also agree with the ATLAS measurements. For \R{2}\ a different
picture is observed. First of all the theoretical predictions change by about $-50$\% when going
from LO to NLO. The origin of this behaviour is traced back to the inclusive two-jet cross section
which is affected by large perturbative corrections. As a function of the leading jet $p_T$, all PDF
sets agree well with the 3/2 ratio ATLAS data at large $p_T$ as shown in figure \Fig{fig:32ratiodata}.
In \Fig{fig:jetratio_pT} we compare LO and NLO predictions for \R{n}\ as function of the leading jet
$p_T$. While for \R{3}\ and \R{4}\ the corrections are moderate for all values of $p_T$ we observe
large negative corrections independent from $p_T$ in case of \R{2}.
Most likely the two-jet rate is very sensitive to soft gluon emission while the higher jet
multiplicities are less affected. As a consequence the fixed-order calculations fail to give
reliable predictions for the 2-jet rate.  A possible improvement could be expected from soft gluon
resummation and matching with parton shower calculations.  As long as only fixed-order calculations
are used to predict \R2\ we do not expect a perfect agreement with the data, especially in the low
$p_T$ region.  Similar to what has been observed in \Fig{fig:MJincl} the comparison with data shows
indeed significant discrepancy in \R2.

\begin{figure}[htbp]
  \begin{center}
    \leavevmode
    \includegraphics[width=1.\columnwidth]{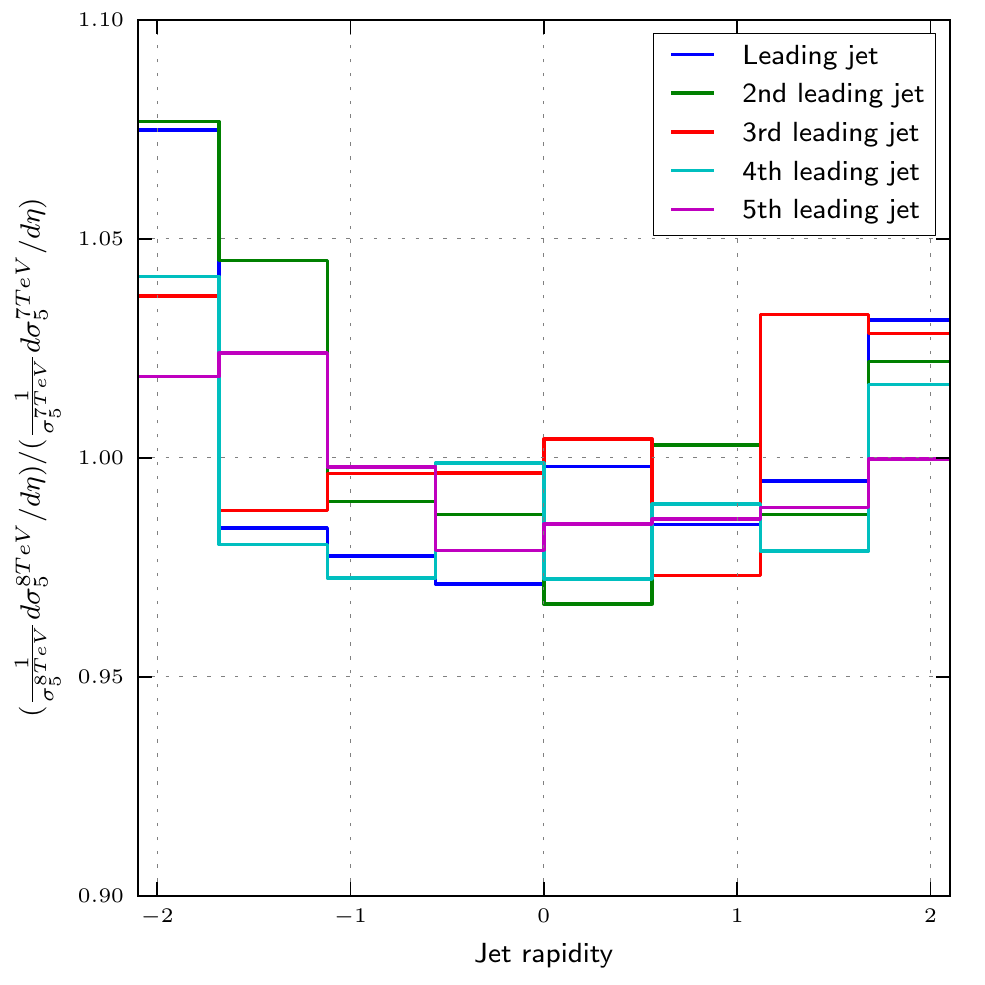}
    \caption{Comparison of LO rapidity distributions of the $p_T$ ordered jets for 7 and 8~TeV}
    \label{fig:7TeVto8TeVratio_eta}
  \end{center}
\end{figure}

\begin{figure}[htbp]
  \begin{center}
    \leavevmode
    \includegraphics[width=1.\columnwidth]{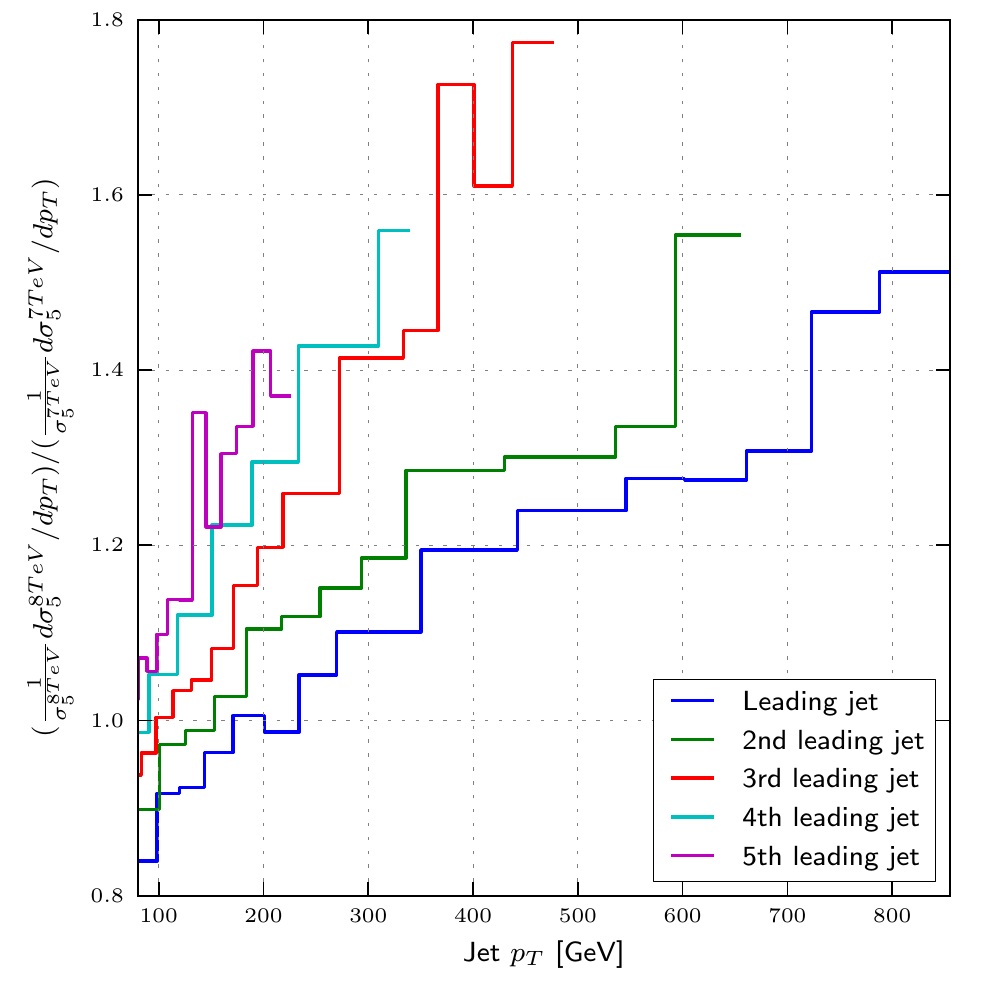}
    \caption{Comparison of LO $p_T$ distributions of the $p_T$ ordered jets for 7 and 8~TeV}
    \label{fig:7TeVto8TeVratio_pt}
  \end{center}
\end{figure}

Let us now move on to less inclusive quantities.
\begin{figure}[htbp]
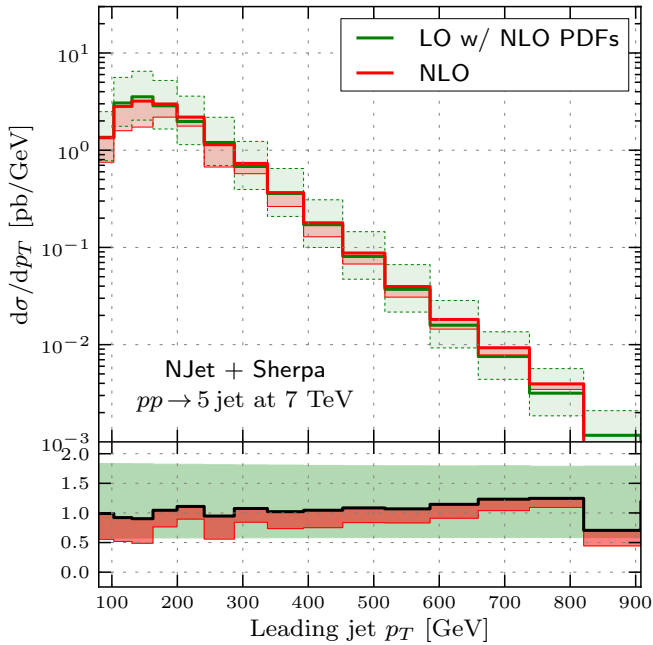

  \begin{center}
    \leavevmode
    \includegraphics[width=1.\columnwidth]{{{plot_q16.4_l8_sm0.1_NNPDF23_7_nlopdf_jet_pT_1}}}
    \caption{The $p_T$ distribution of the leading jet. Both LO and NLO use the NNPDF2.3 PDF set
    with $\alpha_s(M_Z) = 0.118$}
    \label{fig:7TeVpt1dist}
  \end{center}
\end{figure}
In \Fig{fig:7TeVpt1dist} we show the transverse momentum distribution
of the leading jet for five-jet production. Similar to the inclusive
quantities a significant reduction of the scale uncertainty is
observed when going from LO to NLO. Using again the NLO setup to
calculate the LO predictions, the NLO calculation gives very small
corrections. Over a wide range the LO predictions are modified by less
than 10\%. A remarkable feature observed already in the 4-jet
calculation \cite{Bern:2011ep,Badger:2012pf} is the almost constant
K-factor. Again the dynamical scale seems to re-sum possible large
logarithms which would appear at large transverse momentum
using a fixed scale.
Similar findings apply to transverse momentum distribution of the
sub-leading jets.
\begin{figure}[htbp]
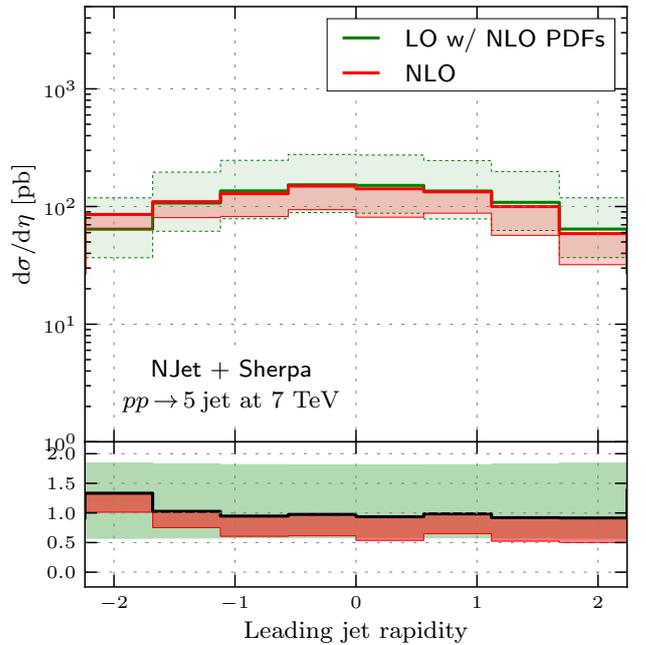

  \begin{center}
    \leavevmode
    \includegraphics[width=1.\columnwidth]{{{plot_q20.4_l10_sm0.1_NNPDF23_7_nlopdf_jet_eta_1}}}
    \caption{The rapidity distribution of the leading jet. Both LO and NLO use the NNPDF2.3 PDF set
    with $\alpha_s(M_Z) = 0.118$}
    \label{fig:7TeVeta1dist}
  \end{center}
\end{figure}
In \Fig{fig:7TeVeta1dist} we show the rapidity distribution of the leading jet, again in LO and NLO
QCD. In the range $-2 < \eta < 2$ the distribution is remarkably flat. Again the NLO corrections are
below 10\% for most $\eta$ values and the K-factor is roughly constant.  We have also investigated
differential distributions for a centre-of-mass energy of 8~TeV. Studying normalized distributions
to account for the increase of the inclusive jet cross section when going from 7 to 8~TeV we
find a remarkable agreement between the 7 and 8~TeV predictions. As example we present in
\Fig{fig:7TeVto8TeVratio_eta} the double-ratio,
\begin{displaymath}
  \left.{1\over \sigma_5^{7\text{TeV-LO}}}{{d\sigma_5^{7\text{TeV-LO}}
  \over d\eta}}\right/
  {1\over \sigma_5^{8\text{TeV-LO}}}{{d\sigma_5^{8\text{TeV-LO}}
  \over d\eta}}.
\end{displaymath}
For simplicity we do not
expand the double ratio in \as. Since the NLO corrections are moderate in size we do not expect a
significant change in the prediction---the difference is formally of
  higher order in \as. As can be see in Fig.~\ref{fig:7TeVto8TeVratio_eta}, the normalized rapidity
distribution changes by less than 5\% when going from 7 to 8~TeV.
For the transverse momentum distribution we expect a harder
  spectrum for 8~TeV centre-of-mass energy compared to 7~TeV. This is
  indeed observed in \Fig{fig:7TeVto8TeVratio_pt}. The fact that for
  low transverse momenta the ratios are below one is an effect of the
  normalization to the total cross section. For 8~TeV the regions where
  the inclusive cross section gets significant contributions is extended
  to larger $p_T$ leading to a ratio below one when comparing with the
  7~TeV case.
Using data for jet production may provide useful input to constrain PDFs. In this context it is very
interesting to study the decomposition of the jet rates with respect to individual partonic channels
not only for inclusive quantities but also for differential distributions.  In
\Fig{fig:channeldecomp-eta} the decomposition of the rapidity distribution of the leading jet is
shown. As in the inclusive case we restrict the discussion to leading order.
\begin{figure}[htbp]
  \begin{center}
    \leavevmode
    \includegraphics[width=1.\columnwidth]{{{plot_channeldecomp_jet_eta_1}}}
    \caption{Contribution of the different partonic channels to the
    leading jet rapidity distribution.}
    \label{fig:channeldecomp-eta}
  \end{center}
\end{figure}
Evidently we find again that the $qg\to q+4g$ channel is the most important channel followed by the
pure gluonic channel. Since the rapidity distribution is only mildly affected by the partonic
centre-of-mass energy we do not expect a strong dependence of the composition with respect to the
rapidity. Indeed as can be seen from \Fig{fig:channeldecomp-eta} the decomposition shows only a weak
dependence on the rapidity. This information can be used to define control samples, when using jet
data to constrain the parton luminosities.
\begin{figure}[htbp]
  \begin{center}
    \leavevmode
    \includegraphics[width=1.\columnwidth]{{{plot_channeldecomp_jet_pT_1}}}
    \caption{Contribution of the different partonic channels to the
    leading jet $p_T$ distribution.}
    \label{fig:channeldecomp-pT}
  \end{center}
\end{figure}
In \Fig{fig:channeldecomp-pT} the analogous results for the transverse momentum distribution is
presented. In difference to the rapidity distribution a significant dependence of the decomposition
as function of the transverse momentum is visible. While at small transverse momentum the $gg\to 5g$
dominates over $qq\to 2q+3g$ the situation changes at about 300 GeV and the $qq\to 2q+3g$ becomes
more important than $gg\to 5g$. This behaviour is a direct consequence of the fact that at high
partonic centre-of-mass energy the quark luminosity ${\cal L}_{q\bar q}$ dominates over the gluon
flux ${\cal L}_{gg}$. A similar pattern, although less pronounced, can also be observed in the
$qq\to 5g$ and $gg\to 4q+1g$ channels.  A cut in the transverse momentum can thus be used to change
the mixture of the individual partonic channels and to provide additional information on specific
parton luminosities.  From the above discussion we expect that different PDF sets should give very
similar results for the rapidity distribution since each bin is rather inclusive with respect to the
partonic centre-of-mass energies where the luminosities are sampled. On the other hand if any
difference using PDF sets from different groups is observed it will most likely show up in the
transverse momentum distribution.
\begin{figure}[htbp]
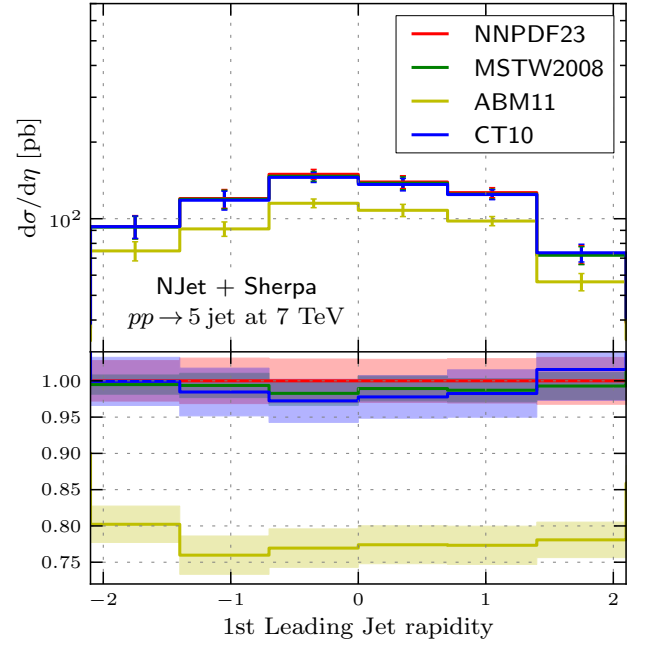

  \begin{center}
    \leavevmode
    \includegraphics[width=1.\columnwidth]{{{plotpdf_as118_HT.50_7_q16.4_l8_jet_eta_1}}}
    \caption{Comparison of different PDF sets on the rapidity distributions. All sets
    use $\alpha_s(M_Z) = 0.118$. The lower plot shows the ratio of the various PDF sets with respect to NNPDF2.3. Error bars in the upper plot are MC errors while shaded areas
    in the lower plot are PDF uncertainties.}
    \label{fig:pdf_eta1}
  \end{center}
\end{figure}
In \Fig{fig:pdf_eta1} the rapidity distribution is shown using four different PDF sets. The PDF sets
NNPDF2.3, CT10 and MSTW2008 lead to very similar results. A major difference is observed comparing
the aforementioned PDF sets with ABM11. ABM11 leads to reduction of about 20\% with respect to
NNPDF2.3, CT10 and MSTW2008. However one can see that the shape for the distribution predicted by
ABM11 agrees well with the other PDF sets. In \Fig{fig:pdf_eta1_norm} and \Fig{fig:pdf_pt1_norm} we
show results for the normalized distributions.
\begin{figure}[htbp]
  \begin{center}
    \leavevmode
    \includegraphics[width=1.\columnwidth]{{{plotpdf_as118_HT.50_7_q16.4_l8_jet_eta_1_norm}}}
    \caption{Comparison of different PDF sets on the rapidity distributions normalized
    to the total cross section. The lower plot shows the ratio of the various PDF sets with respect to NNPDF2.3. Error bars in the upper plot are MC errors while shaded areas
    in the lower plot are PDF uncertainties.}
    \label{fig:pdf_eta1_norm}
  \end{center}
\end{figure}
For the rapidity distribution the four different PDF sets agree well within $\pm 5$\%. The rapidity
distributions of the sub leading jets show a similar behaviour.
\begin{figure}[htbp]
  \begin{center}
    \leavevmode
    \includegraphics[width=1.\columnwidth]{{{plotpdf_as118_HT.50_7_q16.4_l8_jet_pT_1_norm}}}
    \caption{Comparison of different PDF sets on the $p_T$ distributions normalized
    to the total cross section. The lower plot shows the ratio of the various PDF sets with respect to NNPDF2.3. Error bars in the upper plot are MC errors while shaded areas
    in the lower plot are PDF uncertainties.}
    \label{fig:pdf_pt1_norm}
  \end{center}
\end{figure}
In \Fig{fig:pdf_pt1_norm} the transverse momentum distribution is
studied for different PDF sets. As expected one can observe a
  richer structure at large transverse
momentum. While minor differences are visible between ABM11, MSTW08
and NNPDF they still give rather similar results at large $p_T$ at the
level of $10-15$\%. However the CT10 PDF set leads to significantly
larger results at large $p_T$.  In the highest $p_T$ bin the
normalised cross section is enhanced by about $20-30$ \% compared to
ABM11, MSTW08 and NNPDF. Unfortunately it is not easy to distinguish
the different predictions experimentally since the cross section for
this bin is reduced by several orders of magnitude. A significant
amount of data is thus required to achieve the required statistical
sensitivity.

\section{Conclusions}

In this article we have presented first results for five-jet production at NLO accuracy in QCD.
We find moderate corrections at NLO with respect to a leading order computation using NLO PDFs.
Typically corrections of the order of 10\% are observed. Identifying renormalization and factorization scale
and using the total transverse momentum $\HThat$ as dynamical scale leads to a flat K-factor for the
differential distributions. We have compared theoretical predictions
for inclusive jet cross sections and jet
rates with data from ATLAS. With the exception of quantities affected by the two jet rate we find
good agreement between theory and data. As a major uncertainty of the theoretical predictions we
have investigated the impact of using different PDF sets. While for rather inclusive quantities and
distributions not sensitive to a specific partonic centre-of-mass energy good agreement of different sets is
observed (in case of distributions this requires to study normalized predictions) significant
differences are observed in the transverse momentum distribution of the leading jet at large
momentum.

The analysis of the $(n+1)/n$ jet ratios shows that the $4/3$ and $5/4$ predictions appear to be
perturbatively more stable than the $3/2$ predictions with modest correction at NLO. This indicates
that these quantities are good candidates for future extractions of $\alpha_s$ from the LHC
data, where reliable fixed order predictions are mandatory. We hope the results presented
here will be useful in these and other analyses in the future.

\section*{Acknowledgements}

This work is supported by the Helmholtz Gemeinschaft under contract HA-101 (Alliance Physics at the
Terascale), by the German Research Foundation (DFG) through the transregional collaborative research
centre ``Computational Particle Physics'' (SFB-TR9) and by the European Commission through contract
PITN-GA-2010-264564 (LHCPhenoNet). We would also like to thank DESY Zeuthen theory group for
providing computer resources.
\appendix
\section{Differential distributions for sub-leading jets}

In this appendix we present all rapidity and $p_T$ distributions for jets ordered in $p_T$.
A complete set of histograms and plots for $\sqrt{s} = 7$ and $8$ TeV can be obtained from
\url{https://bitbucket.org/njet/njet/wiki/Results/Physics}.

\begin{figure}[h]
  \begin{center}
    \includegraphics[width=1.\columnwidth]{{{plot_q16.4_l8_sm0.1_NNPDF23_7_jet_pT_1}}}
  \end{center}
  \caption{$p_T$ distribution of the leading jet. LO uses NNPDF2.1 with $\alpha_s(M_Z) = 0.119$, NLO
  uses NNPDF2.3 $\alpha_s(M_Z) = 0.118$.}
  \label{fig:pt1}
\end{figure}

\begin{figure}[h]
  \begin{center}
    \includegraphics[width=1.\columnwidth]{{{plot_q16.4_l8_sm0.1_NNPDF23_7_jet_pT_2}}}
  \end{center}
  \caption{$p_T$ distribution of the second leading jet.}
  \label{fig:pt2}
\end{figure}

\begin{figure}[h]
  \begin{center}
    \includegraphics[width=1.\columnwidth]{{{plot_q16.4_l8_sm0.1_NNPDF23_7_jet_pT_3}}}
  \end{center}
  \caption{$p_T$ distribution of the third leading jet.}
  \label{fig:pt3}
\end{figure}

\begin{figure}[h]
  \begin{center}
    \includegraphics[width=1.\columnwidth]{{{plot_q16.4_l8_sm0.1_NNPDF23_7_jet_pT_4}}}
  \end{center}
  \caption{$p_T$ distribution of the fourth leading jet.}
  \label{fig:pt4}
\end{figure}

\begin{figure}[h]
  \begin{center}
    \includegraphics[width=1.\columnwidth]{{{plot_q16.4_l8_sm0.1_NNPDF23_7_jet_pT_5}}}
  \end{center}
  \caption{$p_T$ distribution of the fifth leading jet.}
  \label{fig:pt5}
\end{figure}

\begin{figure}[h]
  \begin{center}
    \includegraphics[width=1.\columnwidth]{{{plot_q20.4_l10_sm0.1_NNPDF23_7_jet_eta_1}}}
  \end{center}
  \caption{$\eta$ distribution of the leading jet. LO uses NNPDF2.1 with $\alpha_s(M_Z) = 0.119$, NLO
  uses NNPDF2.3 $\alpha_s(M_Z) = 0.118$.}
  \label{fig:eta1}
\end{figure}

\begin{figure}[h]
  \begin{center}
    \includegraphics[width=1.\columnwidth]{{{plot_q20.4_l10_sm0.1_NNPDF23_7_jet_eta_2}}}
  \end{center}
  \caption{$\eta$ distribution of the second leading jet.}
  \label{fig:eta2}
\end{figure}

\begin{figure}[h]
  \begin{center}
    \includegraphics[width=1.\columnwidth]{{{plot_q20.4_l10_sm0.1_NNPDF23_7_jet_eta_3}}}
  \end{center}
  \caption{$\eta$ distribution of the third leading jet.}
  \label{fig:eta3}
\end{figure}

\begin{figure}[h]
  \begin{center}
    \includegraphics[width=1.\columnwidth]{{{plot_q20.4_l10_sm0.1_NNPDF23_7_jet_eta_4}}}
  \end{center}
  \caption{$\eta$ distribution of the fourth leading jet.}
  \label{fig:eta4}
\end{figure}

\begin{figure}[h]
  \begin{center}
    \includegraphics[width=1.\columnwidth]{{{plot_q20.4_l10_sm0.1_NNPDF23_7_jet_eta_5}}}
  \end{center}
  \caption{$\eta$ distribution of the fifth leading jet.}
  \label{fig:eta5}
\end{figure}
\FloatBarrier

\section{Numerical results for differential distributions}
\label{sec:tables}
We provide numerical values for the histograms in figures \ref{fig:7TeVpt1dist} and
\ref{fig:7TeVeta1dist} in tables \ref{tab:7TeVpt1dist} and \ref{tab:7TeVeta1dist} for ease of future comparisons.
\FloatBarrier

\begin{table}
  \begin{tabular}{ccc}
    \hline 
    $p_{T_1}$ (GeV) & LO (pb) & NLO (pb) \\
    \hline
   $80.0$ --- $98.4$ & $21.4$ $(0.8)$ & $20.6$ $(6.3)$ \\
   $98.4$ --- $119.7$ & $56.1$ $(1.3)$ & $51.0$ $(7.7)$ \\
   $119.7$ --- $143.9$ & $86.4$ $(1.4)$ & $76.4$ $(7.5)$ \\
   $143.9$ --- $171.0$ & $92.1$ $(1.1)$ & $94.9$ $(5.4)$ \\
   $171.0$ --- $201.1$ & $82.3$ $(1.1)$ & $81.9$ $(4.1)$ \\
   $201.1$ --- $234.0$ & $66.8$ $(1.1)$ & $73.7$ $(4.4)$ \\
   $234.0$ --- $269.8$ & $49.7$ $(0.7)$ & $48.2$ $(3.5)$ \\
   $269.8$ --- $308.5$ & $35.1$ $(0.4)$ & $38.8$ $(2.7)$ \\
   $308.5$ --- $350.2$ & $22.7$ $(0.3)$ & $23.8$ $(1.4)$ \\
   $350.2$ --- $394.7$ & $14.2$ $(0.2)$ & $13.5$ $(1.1)$ \\
   $394.7$ --- $442.2$ & $8.6$ $(0.2)$ & $8.1$ $(0.7)$ \\
   $442.2$ --- $492.5$ & $4.96$ $(0.10)$ & $6.4$ $(1.0)$ \\
   $492.5$ --- $545.8$ & $2.81$ $(0.06)$ & $2.7$ $(0.3)$ \\
   $545.8$ --- $602.0$ & $1.58$ $(0.05)$ & $1.9$ $(0.1)$ \\
   $602.0$ --- $661.1$ & $0.87$ $(0.03)$ & $1.0$ $(0.1)$ \\
   $661.1$ --- $723.0$ & $0.50$ $(0.03)$ & $0.5$ $(0.1)$ \\
   $723.0$ --- $787.9$ & $0.25$ $(0.02)$ & $0.36$ $(0.04)$ \\
   $787.9$ --- $855.7$ & $0.13$ $(0.01)$ & $0.12$ $(0.04)$ \\
   $855.7$ --- $926.4$ & $0.065$ $(0.007)$ & $0.05$ $(0.02)$ \\
   $926.4$ --- $1000.0$ & $0.029$ $(0.005)$ & $0.04$ $(0.01)$ \\
\hline
  \end{tabular}
  \caption{Numerical values for the histograms in figure \ref{fig:7TeVpt1dist}.}\label{tab:7TeVpt1dist}
\end{table}

\begin{table}
  \begin{tabular}{ccc}
    \hline
    $\eta_{1}$ (GeV) & LO (pb) & NLO (pb) \\
    \hline
   $-2.8$ --- $(-2.2)$ & $16.5$ $(1.3)$ & $14.9$ $(4.5)$ \\
   $-2.2$ --- $(-1.7)$ & $36.0$ $(0.9)$ & $47.9$ $(6.4)$ \\
   $-1.7$ --- $(-1.1)$ & $60.0$ $(0.9)$ & $61.6$ $(6.9)$ \\
   $-1.1$ --- $(-0.6)$ & $76.1$ $(0.8)$ & $72.0$ $(4.3)$ \\
   $-0.6$ --- $0.0$ & $85.6$ $(0.8)$ & $83.5$ $(4.3)$ \\
   $0.0$ --- $0.6$ & $84.6$ $(0.8)$ & $79.1$ $(5.4)$ \\
   $0.6$ --- $1.1$ & $75.8$ $(0.8)$ & $74.6$ $(3.6)$ \\
   $1.1$ --- $1.7$ & $60.8$ $(1.0)$ & $55.9$ $(3.7)$ \\
   $1.7$ --- $2.2$ & $36.0$ $(0.8)$ & $33.0$ $(3.8)$ \\
   $2.2$ --- $2.8$ & $15.2$ $(1.1)$ & $21.1$ $(5.7)$ \\
\hline
  \end{tabular}
  \caption{Numerical values for the histograms in figure \ref{fig:7TeVeta1dist}.}\label{tab:7TeVeta1dist}
\end{table}


%
\end{document}